%% file: synergy.tex
\documentclass[12pt, fullpage]{article} 
\input{frontstuff}
\usepackage{mywidefile}

\begin{document} 

\title{On individual neutrality and collective decision making}
\author{Mu Zhu$^{1,*}$, Shangsi Wang$^{1}$, Lu Xin$^{1}$\\[2mm]
{\small $^{1}$ University of Waterloo, 200 University Ave W, Waterloo, ON, Canada N2L 3G1.} \\
{\small $^{*}$ Corresponding author; email: \url{m3zhu@uwaterloo.ca}.}
}
\date{\today}
\maketitle

\input{synergy-main-mathsci}
\input{synergy-proofs}

\subsection*{Acknowledgments}
This work is partially supported by the Natural Sciences and
Engineering Research Council (NSERC) of Canada.

\bibliographystyle{/u/m3zhu/natbib}
\bibliography{/u/m3zhu/mzstat}

\end{document}

%% file: frontstuff.tex
\usepackage{url, float}
\usepackage{amsfonts, amssymb, amscd}
\usepackage{amsmath, delarray, theorem}
\usepackage{natbib}
\bibpunct{(}{)}{;}{a}{}{,}
\renewcommand{\cite}{\citep}

\newcommand{\mycap}[1]{\caption{{\it #1}}}
\newcommand{\bitem}{\begin{itemize}}
\newcommand{\eitem}{\end{itemize}}
\newcommand{\benum}{\begin{enumerate}}
\newcommand{\eenum}{\end{enumerate}}
\newcommand{\beqnn}{\begin{eqnarray*}}
\newcommand{\eeqnn}{\end{eqnarray*}}
\newcommand{\beqn}{\begin{eqnarray}}
\newcommand{\eeqn}{\end{eqnarray}}

{\theoremstyle{plain}
  \newtheorem{proposition}{Proposition}}
{\theoremstyle{plain}
  \newtheorem{corollary}{Corollary}}
{\theoremstyle{plain}
  \newtheorem{theorem}{Theorem}}
{\theoremstyle{plain}
  \newtheorem{lemma}{Lemma}}
{\theoremstyle{plain}
  \newtheorem{definition}{Definition}}

%% file: synergy-main-mathsci.tex

\def\mT{\mathcal{H}}
\def\mF{\mathcal{S}}
\def\t{\mathcal{G}}
\def\f{\mathcal{B}}
\def\auc{V}

\begin{abstract}
We derive a simple mathematical ``theory'' to show that two 
decision-making entities can work better together {\em only if} at least 
one of them is occasionally willing to stay neutral. This provides a
mathematical ``justification'' for an age-old clich\'{e} among marriage 
counselors.
\end{abstract}

{\bf Key words}: average; Bayes theorem; probability; ROC curve; synergy; 
utility. 

{\bf AMS classification}: 62C05.

\section{Introduction}

Suppose that, for each decision one has to make, there are two possible
courses of action, a good one ($\t$) and a bad one ($\f$).
Let $h$ represent a decision-making agent, where
$h(\t) > h(\f)$ means $h$ favors the good action;
$h(\t) = h(\f)$ means $h$ is neutral; and
$h(\t) < h(\f)$ means $h$ favors the bad action.
In addition, suppose that the payoffs associated with
favoring the good action,
being neutral, and favoring the bad action are $+1$, $0$, and $-1$,
respectively.
Let $V(h)$ denote the expected payoff for $h$.
Then,
\beqn
 V(h) =
 (+1) \times P(h(\t)>h(\f)) +
 0    \times P(h(\t)=h(\f)) +
 (-1) \times P(h(\t)<h(\f)). 
\label{eq:EV}
\eeqn
Clearly, $V(h)$ is a measure of the tendency for $h$ to favor the good 
action:
$V(h)>0$ means $h$ is more likely to take the good action;
$V(h)<0$ means it is more likely to take the bad one; and
$V(h)=0$ means it essentially acts randomly.

Let $h_1$ and $h_2$ be two decision-making agents, e.g., a husband and 
wife. In such settings, it is natural to require (e.g., for concerns of 
fairness) that neither $h_1$ nor $h_2$ be allowed to dominate the 
collective decision. This means their respective votes must be on the same 
scale and equally weighted.
Without loss of generality, we assume that both $h_1$
and $h_2$ are binary functions, with
$h_i(x) = 1$ meaning that $h_i$ voted for action $x$ and
$h_i(x) = 0$ meaning that $h_i$ voted against action $x$.
The collective decision is represented by their mean,
\beqnn
 \bar{h}=\frac{h_1+h_2}{2}.
\eeqnn

\begin{definition}
The collective decision $\bar{h}$ is said to be {\em synergistic} if
\beqn
V(\bar{h}) &\geq& \frac{V(h_1)+V(h_2)}{2};
\eeqn
it is said to have {\em positive synergy} if the above inequality
is strict.
\end{definition}

In what follows, we derive ``if and only if'' conditions for the 
collective decision $\bar{h}$ to be synergistic and examine the 
implications of those conditions. Before we proceed, however, it is first
necessary for us to explain why our definition of synergy above 
should depend on the specific payoff structure ($+1, 0, -1$) that gave 
rise to the expression $V(h)$ in equation~(\ref{eq:EV}). 

\subsection{Connection to existing literature: ROC curves}
\label{sec:disc}

Our choice of $V(h)$ is deeply related to the so-called receiver-operating 
characteristic (ROC) curve \citep[][]{rocbk}, a concept widely used in 
many scientific fields such as signal detection and medical diagnostic 
tests.

In the context of medical diagnostic tests, let $h(x)$ be a measurement 
taken on the patient $x$. Suppose the test declares $x$ to be ``sick'' if 
$h(x) < c$ for some threshold $c$, and ``healthy'' if $h(x) \geq c$. The 
test is not perfect. For a given $c$, some healthy patients are declared 
``sick'' (false positives), while some sick patients are declared 
``healthy'' (false negatives).

It is easy to see that decreasing the threshold $c$ will always reduce the
false positive rate but increase the false negative rate of such a test,
while increasing the threshold $c$ will always reduce its false negative 
rate but increase its false positive rate. For example, in the extreme
case of $c = -\infty$, nobody is declared ``sick'' so there can be no
false positives, but the false negative rate reaches 100\% since everyone
is declared ``healthy'' including all truly sick individuals. The
situation is similar but reversed in the other extreme case of $c =
+\infty$.
There is an inherent trade-off between the two types of errors ---
reducing one always increases the other, and the optimal choice
of $c$ depends on their relative costs.
 
The ROC curve essentially traces the two types of errors over the entire
range of $c$, and the area under the ROC curve, or simply ``area under the
curve'' (AUC), is a commonly used performance metric for evaluating these
diagnostic tests.
The AUC has the advantage of not depending on the relative costs of the  
two types of errors. In
addition, it has the following interesting probabilistic interpretation
\citep{auc-prob, rocbk}: Suppose $\mT$ denotes the set of all healthy
patients,
and $\mF$ denotes the set of all sick patients.
Then,
\beqn
\label{eq:AUC}
 \mbox{AUC}(h) \equiv
 1 \times P(h(\t)>h(\f)) +
 \frac{1}{2} \times P(h(\t)=h(\f)) + 
 0 \times P(h(\t)<h(\f))
\eeqn
for any randomly chosen $\t \in \mT$ and $\f \in \mF$.
If $P(h(\t)=h(\f))=0$, meaning $h$ does not produce ties between healthy
and sick patients, this is simply the ``folklore'' statement that
$\mbox{AUC}(h)$ is the probability that the test $h$ correctly orders  
healthy patients ahead of sick ones. The factor 
``1/2'' means   
ties are broken at random.

It is easy to see from (\ref{eq:EV}) and (\ref{eq:AUC}) that
\[ 
 V(h) = 2\mbox{AUC}(h) - 1.
\]
In other words, our expected payoff $V(h)$, as defined in
(\ref{eq:EV}), is merely a linearly transformed version of the
widely-used AUC for evaluating the effectiveness of medical diagnostic
tests. One can think of
$V(h)$ as measuring the ability of $h$ to ``diagnose'' the
good action when faced with a decision.

\section{Results}

We now state two lemmas for respectively the independent and 
dependent cases. Together, they establish necessary and sufficient 
conditions for the collective decision $\bar{h}$ to be synergistic. Proofs 
of the two lemmas are given in the appendices. However, the punchline of 
the paper, which we state in Section~\ref{sec:punchline}, is a consequence 
of these two lemmas, rather than the two lemmas themselves.

\subsection{Independent case}

First, suppose $h_1$ and $h_2$ act independently.
For $i=1,2$, let
\beqnn
a_i &=& P(h_i(\t) > h_i(\f)), \\
b_i &=& P(h_i(\t) = h_i(\f)), \\
c_i &=& P(h_i(\t) < h_i(\f)).
\eeqnn
In other words, 
$a_i$ 
is the probability that $h_i$ favors the good action; $b_i$ is the 
probability that $h_i$ is neutral; and $c_i$ is the 
probability that $h_i$ favors the bad action.  

\begin{lemma}
\label{lemma:auc}
Suppose $h_1$ and $h_2$ are statistically independent. 
Then, $\bar{h}$ is synergistic
if and only if
\beqn
\label{eq:iff-indep}
  b_1 (a_2 - c_2) + b_2 (a_1 - c_1) \geq 0.
\eeqn
\end{lemma}

\begin{corollary}
For $\bar{h}$ to be synergistic, it is sufficient (but not necessary) for
both $a_1 \geq c_1$ and $a_2 \geq c_2$.
\end{corollary}
Since $V(h_1) = a_1 - c_1$ by (\ref{eq:EV}) and likewise for $h_2$, the 
conclusion here is simply this: the collective decision is synergistic as 
long as both decision makers are ``no worse than random'', having 
a slightly higher chance of favoring the good action rather than the bad 
one. 

\subsubsection{Connection to existing literature: Weak learners}

The notion of being ``no worse than random'' is analogous to that of a 
``weak learner'' \citep{weak-learner}, an important concept in PAC 
learning theory \citep{valient84} and the theory of boosting 
\citep[][]{boosting-orig}.

\subsection{Dependent case}

More generally, suppose that $h_1$ and $h_2$ are not independent.
For clarity, we make a small change in the notation to describe this case, 
using lowercase letters $a,b,c$ for the 
marginal probabilities of $h_1$, and uppercase letters $A,B,C$ for those 
of $h_2$:
\beqnn
&a = P(h_1(\t) > h_1(\f)), &\quad A = P(h_2(\t) > h_2(\f)); \\
&b = P(h_1(\t) = h_1(\f)), &\quad B = P(h_2(\t) = h_2(\f)); \\
&c = P(h_1(\t) < h_1(\f)), &\quad C = P(h_2(\t) < h_2(\f)). 
\eeqnn
Furthermore, since $h_1$ and $h_2$ are dependent, we use $P_{xX}$ to refer 
to the conditional probabilities of $h_2$ given $h_1$, and $Q_{Yy}$ to 
refer to the conditional probabilities of $h_1$ given $h_2$,
as displayed 
in Table~\ref{tab:PQ}. 
These notations for $P$ and $Q$ are 
analogous to those typically used to denote transition probabilities 
for Markov chains \citep{ross}.  

Of course, $P$ and $Q$ are related by Bayes 
theorem \citep{bayes}. For example,
\begin{multline}
P(h_2(\t)=h_2(\f)|h_1(\t)<h_1(\f)) 
= ... \\
= \frac{P(h_1(\t)<h_1(\f)|h_2(\t)=h_2(\f)) \times P(h_2(\t)=h_2(\f))}
     {P(h_1(\t)<h_1(\f))}
\end{multline}
or, using the compact notations of Table~\ref{tab:PQ},
\beqnn
P_{cB} = \frac{Q_{Bc} \times B}{c},
\quad\mbox{or}\quad
c P_{cB} = B Q_{Bc}.
\eeqnn
This leads to the following proposition, which is needed in order to prove 
Lemma~\ref{lemma:auc-nonindep} that follows.

\begin{proposition}
\label{prop:bayes}
For $x = a, b, c$ and $Y = A, B, C$, $x P_{xY} = Y Q_{Yx}$.
\end{proposition}

\begin{lemma}
\label{lemma:auc-nonindep}
Suppose $h_1$ and $h_2$ are statistically dependent, with conditional 
probabilities given by Table~\ref{tab:PQ}. 
Then, $\bar{h}$ is synergistic
if and only if
\beqn
\label{eq:iff-nonindep}
 b \left(P_{bA} - P_{bC}\right) +
 B \left(Q_{Ba} - Q_{Bc}\right) \geq 0.
\eeqn
\end{lemma}

\begin{corollary}
For $\bar{h}$ to be synergistic, it is sufficient (but not necessary) for
both
$P_{bA} \geq P_{bC}$
and
$Q_{Ba} \geq Q_{Bc}$. 
\end{corollary}
Notice the similarity and symmetry of the two ``if and only if'' 
conditions, (\ref{eq:iff-indep}) and (\ref{eq:iff-nonindep}). They can 
both be interpreted as follows: ``Whenever one decision maker is neutral, 
the other one is more likely to favor the good action rather than the bad 
one.'' Clearly, this is a highly intuitive characterization of the notion of 
synergy.

\begin{table}[ht]
\centering
\mycap{\label{tab:PQ}%
Conditional probabilities.
(I) $P_{xX}$ denotes various conditional probabilities of $h_2$ given
$h_1$.
(II) $Q_{Yy}$ denotes various conditional probabilities of $h_1$ given
$h_2$. 
For example,
$P_{aB} = P(h_2(\t)=h_2(\f)|h_1(\t)>h_1(\f))$,
$Q_{Bc} = P(h_1(\t)<h_1(\f)|h_2(\t)=h_2(\f))$, etc.}
(I) \\[2mm]
\fbox{%
\begin{tabular}{l|ccc}   
 &         $h_2(\t) > h_2(\f)$ & $h_2(\t) = h_2(\f)$ & $h_2(\t) < h_2(\f)$ 
\\
\hline 
$h_1(\t) > h_1(\f)$
 & $P_{aA}$ & $P_{aB}$ & $P_{aC}$ \\
$h_1(\t) = h_1(\f)$
 & $P_{bA}$ & $P_{bB}$ & $P_{bC}$ \\
$h_1(\t) < h_1(\f)$   
 & $P_{cA}$ & $P_{cB}$ & $P_{cC}$ \\
\end{tabular}}\\[2mm]
(II) \\[2mm]
\fbox{%
\begin{tabular}{l|ccc}
 &         $h_1(\t) > h_1(\f)$ & $h_1(\t) = h_1(\f)$ & $h_1(\t) < h_1(\f)$ 
\\
\hline
$h_2(\t) > h_2(\f)$
 & $Q_{Aa}$ & $Q_{Ab}$ & $Q_{Ac}$ \\
$h_2(\t) = h_2(\f)$
 & $Q_{Ba}$ & $Q_{Bb}$ & $Q_{Bc}$ \\
$h_2(\t) < h_2(\f)$
 & $Q_{Ca}$ & $Q_{Cb}$ & $Q_{Cc}$ \\
\end{tabular}}
\end{table}

\section{Conclusion}
\label{sec:punchline}

It is easy to see that the inequality (\ref{eq:iff-indep}) will become an 
equality when $b_1=b_2=0$. The same can be said for inequality 
(\ref{eq:iff-nonindep}) when $b=B=0$. These cases correspond to the 
situation where neither $h_1$ nor $h_2$ is ever neutral about a decision.

\begin{definition}
A decision-making agent $h$ is said to be {\em opinion-loaded}
if its probability of being neutral is zero, i.e.,
if $P(h(\t)=h(\f))=0$.
\end{definition}

\begin{theorem}
There can be no positive synergy in the collective decision $\bar{h}$ if 
both decision-making agents are opinion-loaded.
\end{theorem}

Our results, therefore, imply that there can be no positive synergy in the 
collective decision without individual neutrality. In other words, the 
willingness to compromise is not an option; it is a necessity!

Intuitively, this is because opinion-loaded decision-makers are never 
willing to admit that they may sometimes have a hard time making a good 
decision. As a result, they don't give the other decision-maker a chance 
to take over the decision when it can be beneficial to do so; this 
explains why there can be no positive synergy. In order for there to be 
positive synergy at all, it is necessary for at least one agent to have a 
strictly positive probability of remaining neutral.

For centuries, marriage counselors must have been giving such advice to 
couples all over the world, but they may not be aware of the mathematical 
justification for their age-old practice.

%% file: synergy-proofs.tex
\appendix

\section{Proof of Lemma~\ref{lemma:auc}}

Since 
both $h_1$ and $h_2$ are binary functions, $\bar{h}(\t)>\bar{h}(\f)$ if 
and only if
\bitem
\item[(i)] 
$h_1(\t)>h_1(\f)$ and 
$h_2(\t)>h_2(\f)$; or
\item[(ii)] 
$h_1(\t)>h_1(\f)$ and 
$h_2(\t)=h_2(\f)$; or
\item[(iii)]
$h_1(\t)=h_1(\f)$ and
$h_2(\t)>h_2(\f)$. 
\eitem
That $h_1$ and $h_2$ are statistically independent means
\beqn
\label{eq:proof-win}
 P(\bar{h}(\t)>\bar{h}(\f)) = 
a_1 a_2 + a_1 b_2 + a_2 b_1.
\eeqn
Likewise,
\beqn
\label{eq:proof-lose}
 P(\bar{h}(\t)<\bar{h}(\f)) = 
c_1 c_2 + c_1 b_2 + c_2 b_1.
\eeqn
Using the definition (\ref{eq:EV}) and 
(\ref{eq:proof-win})-(\ref{eq:proof-lose}), we get
\beqn
& &
\auc(\bar{h}) - \frac{\auc(h_1)+\auc(h_2)}{2} \notag \\
&=& 
\left[(a_1 a_2 + a_1 b_2 + a_2 b_1) -
      (c_1 c_2 + c_1 b_2 + c_2 b_1)
\right] -
\frac{(a_1-c_1)+(a_2-c_2)}{2} \notag \\
&=&
a_1 a_2 - c_1 c_2 + 
\left(b_2 - \frac{1}{2}\right)(a_1 - c_1) +
\left(b_1 - \frac{1}{2}\right)(a_2 - c_2). \label{eq:proof-stop1}
\eeqn
Since $a_i+b_i+c_i=1$ for $i=1,2$, (\ref{eq:proof-stop1}) becomes
\beqn
& &
a_1 a_2 - c_1 c_2 + 
\left(\frac{1}{2}-a_2-c_2\right)(a_1 - c_1) +
\left(\frac{1}{2}-a_1-c_1\right)(a_2 - c_2) \notag \\
&=&
(a_1 a_2 - c_1 c_2) + 
\frac{a_2-c_2}{2}-(a_2+c_2)(a_1 - c_1) +
\frac{a_1-c_1}{2}-(a_1+c_1)(a_2 - c_2) \notag \\
&=&
\frac{a_2-c_2}{2} +
\frac{a_1-c_1}{2} -
(a_1 a_2 - c_1 c_2) \notag \\ 
&=&
\frac{1}{2} 
 \underbrace{%
 \left[ (a_2-c_2) + (a_1-c_1) - 2 (a_1 a_2 - c_1 c_2) 
 \right]}_{(*)}, \notag 
\eeqn
where
\beqn
(*) &=& 
\left[(a_2 - c_2) - (a_1 a_2 - c_1 c_2)\right] + 
\left[(a_1 - c_1) - (a_1 a_2 - c_1 c_2)\right] \notag \\
&=& 
\left[(1-a_1)a_2 - (1-c_1)c_2 \right] +
\left[(1-a_2)a_1 - (1-c_2)c_1 \right]. \label{eq:proof-stop2} 
\eeqn
Finally, using the fact that $a_i+b_i+c_i=1$ for $i=1,2$ again, we see 
that (\ref{eq:proof-stop2}) 
is equal to
\beqn 
& & 
\left[(b_1+c_1)a_2 - (a_1+b_1)c_2 \right] +
\left[(b_2+c_2)a_1 - (a_2+b_2)c_1 \right] \notag \\
&=& 
b_1 a_2 - b_1 c_2 +
b_2 a_1 - b_2 c_1 \notag \\
&=& 
b_1 (a_2 - c_2) +
b_2 (a_1 - c_1), \notag 
\eeqn
which proves the lemma. $\square$


\section{Proof of Lemma~\ref{lemma:auc-nonindep}}

By the same argument used to start the proof of Lemma~\ref{lemma:auc},
\beqn
& &
\auc(\bar{h}) - \frac{\auc(h_1)+\auc(h_2)}{2} \notag \\
&=& 
\left[(a P_{aA} + a P_{aB} + b P_{bA}) -
      (c P_{cC} + c P_{cB} + b P_{bC})
\right] -
\frac{(a-c)+(A-C)}{2}. \label{eq:auc2-start}
\eeqn
But 
\beqnn
 A = aP_{aA} + bP_{bA} + cP_{cA} \quad\mbox{and}\quad
 C = aP_{aC} + bP_{bC} + cP_{cC},
\eeqnn
so (\ref{eq:auc2-start}) is equal to
\begin{multline*}
(a P_{aA} + a P_{aB} + b P_{bA}) -
(c P_{cC} + c P_{cB} + b P_{bC}) - ... \\
... -
\frac{(a-c)}{2} -
\frac{(aP_{aA} + bP_{bA} + cP_{cA})-
      (aP_{aC} + bP_{bC} + cP_{cC})}{2}.
\end{multline*}
Gathering terms multiplying $a$, $b$, and $c$, respectively, this becomes
\begin{multline}
a\left(P_{aA}+P_{aB}-\frac{1+P_{aA}-P_{aC}}{2}\right) +
b\left(P_{bA}-P_{bC}-\frac{P_{bA}-P_{bC}}{2}\right) - ...\\
... - 
c\left(P_{cC}+P_{cB}-\frac{1-P_{cA}+P_{cC}}{2}\right). 
\label{eq:auc2-step1}
\end{multline}
Using the fact that
\beqnn
P_{aA} + P_{aB} + P_{aC} = 1 \quad\mbox{and}\quad
P_{cC} + P_{cB} + P_{cC} = 1,
\eeqnn
expression (\ref{eq:auc2-step1}) can be simplified to
\beqn
\label{eq:auc2-almost}
a \frac{P_{aB}}{2} +
b \left( \frac{P_{bA}-P_{bC}}{2} \right) - 
c \frac{P_{cB}}{2}.
\eeqn
However, by Proposition~\ref{prop:bayes}, we have
\[
 a P_{aB} = BQ_{Ba}
\quad\mbox{and}\quad
 c P_{cB} = BQ_{Bc}.
\]
Substituting this into (\ref{eq:auc2-almost}) leads to the conclusion that
(\ref{eq:auc2-start})$\geq 0$ if and only if
\[
 b \left(P_{bA} - P_{bC}\right) +
 B \left(Q_{Ba} - Q_{Bc}\right) \geq 0. \quad\square
\]
